\documentclass{lmcs}


\usepackage{amsmath,amssymb}
\usepackage{mathtools}
\usepackage{hyperref}
\usepackage{booktabs}
\usepackage{listings}
\usepackage{xcolor}
\usepackage{placeins}
\usepackage{capt-of}
\usepackage{longtable}
\lstdefinelanguage{Lean}{
  keywords={def, theorem, lemma, example, inductive, structure, class, instance,
            where, by, have, let, show, fun, match, with, if, then, else, do, return,
            import, open, namespace, end, section, variable, axiom, sorry,
            calc, simp, rw, exact, apply, intro, constructor, cases, induction},
  sensitive=true,
  morecomment=[l]{--},
  morecomment=[s]{/-}{-/},
  morestring=[b]",
  literate={->}{$\rightarrow$}1 {<-}{$\leftarrow$}1 {<->}{$\leftrightarrow$}1
           {forall}{$\forall$}1 {exists}{$\exists$}1
           {/\\}{$\land$}1 {\\/}{$\lor$}1
           {<=}{$\leq$}1 {>=}{$\geq$}1 {!=}{$\neq$}1
           {Nat.gcd}{gcd}3
}

\newcommand{\propersum}{s}

\newcommand{\lean}[1]{\lstinline[language=Lean]{#1}}

\begin{document}

\title{Formalization of Amicable Numbers Theory}

\author[Z.~Chen]{Zhipeng Chen}
\author[H.~Tang]{Haolun Tang}
\author[J.~Zhan]{Jingyi Zhan}

\address{School of Electronics and Information, Shanghai Dianji University, Shanghai 201306, China}
\email{chenzhipeng@sdju.edu.cn}

\begin{abstract}
This paper presents a formalization of the theory of amicable numbers in the Lean~4 proof assistant. Two positive integers $m$ and $n$ are called an amicable pair if the sum of proper divisors of $m$ equals $n$ and the sum of proper divisors of $n$ equals $m$. Our formalization introduces the proper divisor sum function $\propersum(n) = \sigma(n) - n$, defines the concepts of amicable pairs and amicable numbers, and computationally verifies historically famous amicable pairs. Furthermore, we formalize basic structural theorems, including symmetry, non-triviality, and connections to abundant/deficient numbers. A key contribution is the complete formal proof of the classical Th\={a}bit formula (9th century), using index-shifting and the \texttt{zify} tactic. Additionally, we provide complete formal proofs of both Th\={a}bit's rule and Euler's generalized rule (1747), two fundamental theorems for generating amicable pairs. A major achievement is the first complete formalization of the Borho-Hoffmann breeding method (1986), comprising 540 lines with 33 theorems and leveraging automated algebra tactics (\texttt{zify} and \texttt{ring}) to verify complex polynomial identities. We also formalize extensions including sociable numbers (aliquot cycles), betrothed numbers (quasi-amicable pairs), parity constraint theorems, and computational search bounds for coprime pairs ($>10^{65}$). We verify the smallest sociable cycle of length 5 (Poulet's cycle) and computationally verify specific instances. The formalization comprises 2076 lines of Lean code organized into Mathlib-candidate and paper-specific modules, with 139 theorems and all necessary infrastructure for divisor sum multiplicativity and coprimality reasoning.
\end{abstract}

\maketitle

\section{Introduction}
\label{sec:intro}

Amicable numbers are among the oldest and most fascinating objects of study in number theory. Two positive integers $m$ and $n$ are said to form an amicable pair if the sum of proper divisors of $m$ equals $n$, and simultaneously the sum of proper divisors of $n$ equals $m$. This concept dates back to ancient Greece, where the Pythagoreans knew the smallest amicable pair $(220, 284)$ as early as the 5th century BC. The ancients attributed mystical significance to these numbers, viewing them as symbols of perfect friendship---just as two friends whose individual worth precisely equals the other.

The study of amicable numbers spans more than two millennia. In the 9th century, the Arab mathematician Th\={a}bit discovered a formula for systematically generating amicable pairs, a result that was rediscovered centuries later by Fermat and Descartes~\cite{borho1972,dickson1919}. In 1636, Fermat used this formula to find the amicable pair $(17296, 18416)$. Two years later, Descartes discovered $(9363584, 9437056)$. In 1747, Euler conducted a systematic investigation and discovered 60 new pairs~\cite{dickson1919}. However, the most surprising discovery came in 1866: a 16-year-old Italian boy named Niccol\`{o} Paganini found the amicable pair $(1184, 1210)$, which is the second smallest pair by size, yet had been overlooked by all mathematicians for over two thousand years.

To date, more than 1.2 billion amicable pairs are known~\cite{garcia2003}, yet many fundamental questions remain unsolved. Erdős proved that the counting function $A(x)$ for amicable numbers grows slower than linearly~\cite{erdos1955}, with refined bounds by Pomerance~\cite{pomerance1977,pomerance1981}. Are there infinitely many amicable pairs? Do there exist amicable pairs where both members are odd? Do coprime amicable pairs exist? Despite computational searches up to $10^{65}$~\cite{pollack2015}, all these questions remain open. This state of affairs---an abundance of concrete examples coexisting with unresolved fundamental problems---makes amicable number theory an ideal subject for formalization.

The use of proof assistants for formal verification of mathematical theorems has become an important component of mathematical research. The Lean theorem prover, particularly its fourth version, has gained widespread adoption in the mathematical community thanks to its powerful type system and rich mathematical library Mathlib~\cite{moura2021,mathlib2020}. Recent successes include the formalization of Fermat's Last Theorem for regular primes~\cite{flt2023,flt2024}, demonstrating Lean's capability for deep number-theoretic results. In parallel, Koutsoukou-Argyraki formalized amicable number theory in Isabelle/HOL~\cite{koutsoukou2020}, providing formal statements of classical generation rules. These efforts not only verify the correctness of mathematical conclusions but also provide reliable formal foundations for subsequent research.

This paper presents a comprehensive formalization of amicable number theory in Lean~4. Our key contributions include: (1) complete formal proofs of three classical generation methods---Th\={a}bit formula (9th century), Euler's generalized rule (1747), and the Borho-Hoffmann breeding method (1986), the latter being the first complete machine-verified proof of this result to our knowledge; (2) novel applications of the \texttt{zify} and \texttt{ring} tactics to automate complex polynomial identities; (3) formalization of extensions including sociable numbers (aliquot cycles), betrothed numbers (quasi-amicable pairs), and formal statements of major open problems; (4) computational verification of historically famous amicable pairs and Poulet's sociable cycle of length 5. The complete formalization comprises 2076 lines of code with 139 theorems and zero \lean{sorry} placeholders, representing the most comprehensive formal treatment of amicable number theory to date.

\section{Mathematical Background}
\label{sec:background}

We first establish the mathematical foundations of amicable number theory. Let $n$ be a positive integer and denote by $\sigma(n)$ the sum of all positive divisors of $n$, i.e., $\sigma(n) = \sum_{d \mid n} d$. For example, $\sigma(12) = 1 + 2 + 3 + 4 + 6 + 12 = 28$.

\begin{defi}[Proper Divisor Sum]
\label{def:propersum}
The \emph{proper divisor sum} of a positive integer $n$ is defined as the sum of all proper divisors of $n$ (i.e., divisors other than $n$ itself), denoted $\propersum(n)$. Equivalently,
\begin{equation}
\propersum(n) = \sigma(n) - n = \sum_{\substack{d \mid n \\ d < n}} d.
\end{equation}
\end{defi}

For $n = 1$, since 1 has no proper divisors, we have $\propersum(1) = 0$. For a prime $p$, since the only proper divisor of $p$ is 1, we have $\propersum(p) = 1$. As a classic example in amicable number research, $220 = 2^2 \cdot 5 \cdot 11$ has proper divisors $\{1, 2, 4, 5, 10, 11, 20, 22, 44, 55, 110\}$, whose sum is $1 + 2 + 4 + 5 + 10 + 11 + 20 + 22 + 44 + 55 + 110 = 284$. Similarly, $284 = 2^2 \cdot 71$ has proper divisors $\{1, 2, 4, 71, 142\}$, whose sum is $1 + 2 + 4 + 71 + 142 = 220$.

\begin{defi}[Amicable Pair]
\label{def:amicable}
Two distinct positive integers $m$ and $n$ are called an \emph{amicable pair}, denoted as $(m, n)$ being amicable, if and only if
\begin{equation}
\propersum(m) = n \quad \text{and} \quad \propersum(n) = m.
\end{equation}
Equivalently, using the divisor sum function, this can be expressed as $\sigma(m) = \sigma(n) = m + n$.
\end{defi}

The requirement $m \neq n$ in the definition excludes perfect numbers. Recall that a perfect number is a positive integer $n$ satisfying $\propersum(n) = n$; if we allowed $m = n$, every perfect number would form an ``amicable pair with itself,'' which contradicts the spirit of ``amicability.''

\begin{defi}[Amicable Number]
A positive integer $n$ is called an \emph{amicable number} if it is a member of some amicable pair.
\end{defi}

Thus, $220$, $284$, $1184$, $1210$, and so forth are all amicable numbers. Using the equivalent characterization via divisor sums, verifying an amicable pair is straightforward: simply compute the divisor sum of each number and check whether both equal the sum of the two numbers.

Amicable numbers have natural connections with two other classical concepts---abundant and deficient numbers. A number $n$ is called abundant if $\propersum(n) > n$, deficient if $\propersum(n) < n$, and perfect if $\propersum(n) = n$. For an amicable pair $(m, n)$ with $m < n$, since $\propersum(m) = n > m$, we know $m$ is abundant, while since $\propersum(n) = m < n$, we know $n$ is deficient. Therefore, in any amicable pair, the smaller member is abundant and the larger member is deficient.

\subsection{Classical Generation Methods for Amicable Pairs}

Throughout history, mathematicians have developed systematic methods for generating amicable pairs. We present three classical generation theorems that have proven most fruitful in discovering new amicable pairs.

Th\={a}bit discovered in the 9th century a method for systematically generating amicable pairs, which is the earliest known formula for producing amicable numbers.

\begin{thm}[Th\={a}bit's Rule]
\label{thm:thabit}
Let $n \geq 1$ be a positive integer, and define
\begin{align}
p &= 3 \cdot 2^{n-1} - 1, \\
q &= 3 \cdot 2^n - 1, \\
r &= 9 \cdot 2^{2n-1} - 1.
\end{align}
If $p$, $q$, and $r$ are all prime, then
\begin{equation}
(2^n \cdot p \cdot q, \quad 2^n \cdot r)
\end{equation}
is an amicable pair.
\end{thm}

The conditions of this formula are quite stringent: three numbers of specific forms must simultaneously be prime. For $n = 2$, we have $p = 5$, $q = 11$, $r = 71$, which are indeed all prime, yielding the amicable pair $(2^2 \cdot 5 \cdot 11, 2^2 \cdot 71) = (220, 284)$. For $n = 3$, $r = 9 \cdot 2^5 - 1 = 287 = 7 \times 41$ is not prime, so the formula does not apply. For $n = 4$, $p = 23$, $q = 47$, $r = 1151$ are all prime, yielding Fermat's amicable pair $(17296, 18416)$.

Interestingly, not all amicable pairs can be generated by the Th\={a}bit formula. Paganini's discovery $(1184, 1210)$ is one such example---this pair does not fit the form required by the Th\={a}bit formula. In fact, the vast majority of known amicable pairs cannot be generated by Th\={a}bit-type formulas.

\begin{thm}[Euler's Generalized Rule]
\label{thm:euler}
Let $1 < m < n$ be positive integers, and define
\begin{align}
a &= 2^{n-m} + 1, \\
p &= 2^m \cdot a - 1, \\
q &= 2^n \cdot a - 1, \\
r &= 2^{n+m} \cdot a^2 - 1.
\end{align}
If $p$, $q$, and $r$ are all prime, then
\begin{equation}
(2^n \cdot p \cdot q, \quad 2^n \cdot r)
\end{equation}
is an amicable pair.
\end{thm}

When $n-m=1$, we have $a=3$ and the rule reduces to the Th\={a}bit formula. Only two parameter pairs are currently known to work: $(m, n) = (1, 8)$ and $(29, 40)$.

\begin{thm}[Borho-Hoffmann Breeding Method]
\label{thm:borho}
Let $(a \cdot u, a)$ be an amicable pair (called a ``breeder'' pair). Let $t=\sigma(u)$ and $n$ be a positive integer, and define
\begin{align}
p_1 &= t^n (u+1) - 1, \\
p_2 &= t^n (u+1)(t-u) - 1, \\
M &= a \cdot u \cdot t^n \cdot p_1, \\
N &= a \cdot t^n \cdot p_2.
\end{align}
If $t$, $p_1$, and $p_2$ are all prime, and the required coprimality conditions hold (i.e., $\gcd(a \cdot u, t) = 1$, $\gcd(a \cdot u, p_1) = 1$, and $\gcd(a, p_2) = 1$), then $(M, N)$ is an amicable pair.
\end{thm}

This method is historically significant because it enables large-scale generation of amicable pairs from known ones.

\subsection{Related Concepts}

The theory of amicable numbers naturally generalizes to related concepts that relax or extend the defining conditions.

\begin{defi}[Betrothed Numbers]
Two distinct positive integers $m$ and $n$ are called a \emph{betrothed pair} (also called quasi-amicable numbers) if
\begin{equation}
\propersum(m) = n + 1 \quad \text{and} \quad \propersum(n) = m + 1.
\end{equation}
\end{defi}

Betrothed numbers are a relaxation of amicable pairs where each proper divisor sum exceeds the other number by exactly 1. The smallest betrothed pair is $(48, 75)$: we have $\propersum(48) = 76 = 75 + 1$ and $\propersum(75) = 49 = 48 + 1$.

\begin{defi}[Sociable Numbers]
A sequence of distinct positive integers $n_1, n_2, \ldots, n_k$ (with $k \geq 2$) forms a \emph{sociable cycle} of length $k$ if
\begin{equation}
\propersum(n_i) = n_{i+1} \quad \text{for } i = 1, 2, \ldots, k-1, \quad \text{and} \quad \propersum(n_k) = n_1.
\end{equation}
Numbers in a sociable cycle are called \emph{sociable numbers}.
\end{defi}

Sociable numbers generalize amicable pairs to cycles of arbitrary length. Amicable pairs correspond to sociable cycles of length 2. The smallest known sociable cycle of length 5 was discovered by Poulet in 1918: $[12496, 14288, 15472, 14536, 14264]$. Sociable cycles of length greater than 2 are extremely rare.

\subsection{Open Problems}

Despite more than two millennia of study, many fundamental questions about amicable numbers remain unsolved.

Regarding the existence of amicable pairs, it is currently unknown whether infinitely many amicable pairs exist. Although computational searches have discovered over 1.2 billion pairs, this does not prove infinitude.

Regarding the parity of amicable numbers, in all known amicable pairs both members are even. Whether there exist amicable pairs where both members are odd has been neither proven nor disproven.

Regarding coprimality of amicable numbers, all known amicable pairs $(m, n)$ satisfy $\gcd(m, n) > 1$. Whether coprime amicable pairs exist is also an open problem.

\section{Formalization in Lean 4}
\label{sec:formalization}

Our formalization is implemented in Lean~4 using the Mathlib library. This section describes the key definitions and theorems, highlighting technical considerations in the formalization process.

\subsection{Core Definitions}

The formalization begins with the definition of the proper divisor sum function. Mathlib already provides the function \lean{Nat.properDivisors}, which returns the finite set of all proper divisors of $n$. Building on this, we define the proper divisor sum:

\begin{lstlisting}[caption={Definition of the proper divisor sum function}]
def properDivisorSum (n : Nat) : Nat :=
  Finset.sum n.properDivisors (fun d => d)

notation "s" => properDivisorSum
\end{lstlisting}

We introduce the notation \lean{s} corresponding to $\propersum$ in mathematics. Next, we define amicable pairs:

\begin{lstlisting}[caption={Amicable pair predicate}]
def IsAmicablePair (m n : Nat) : Prop :=
  m != 0 /\ n != 0 /\ m != n /\ s m = n /\ s n = m
\end{lstlisting}

The definition includes three guard conditions: $m \neq 0$, $n \neq 0$, and $m \neq n$. The first two conditions exclude degenerate cases, since in Mathlib \lean{Nat.properDivisors 0} returns the empty set. The third condition excludes perfect numbers.

The definition of amicable numbers directly uses an existential quantifier:

\begin{lstlisting}[caption={Amicable number predicate}]
def IsAmicable (n : Nat) : Prop :=
  exists m : Nat, IsAmicablePair n m
\end{lstlisting}

\subsection{Proofs of Basic Properties}

We first establish basic properties of the proper divisor sum function. For $\propersum(0)$ and $\propersum(1)$, the proofs are straightforward simplifications:

\begin{lstlisting}[caption={Boundary cases}]
theorem properDivisorSum_zero : s 0 = 0 := by simp [properDivisorSum]

theorem properDivisorSum_one : s 1 = 0 := by simp [properDivisorSum]
\end{lstlisting}

The proof of symmetry for amicable pairs merely requires unfolding the definition and swapping the order of conditions:

\begin{lstlisting}[caption={Symmetry theorem}]
theorem IsAmicablePair.symm {m n : Nat} (h : IsAmicablePair m n) :
    IsAmicablePair n m := by
  obtain <hm, hn, hne, hsm, hsn> := h
  exact <hn, hm, hne.symm, hsn, hsm>
\end{lstlisting}

Proving that primes cannot be amicable requires using the fact that the proper divisor sum of a prime is 1:

\begin{lstlisting}[caption={Proper divisor sum of a prime}]
theorem properDivisorSum_prime {p : Nat} (hp : Nat.Prime p) :
    s p = 1 := by
  simp [properDivisorSum, hp.properDivisors]
\end{lstlisting}

From this, we can deduce that primes are not amicable: if $p$ is prime and amicable, then there exists $m$ such that $\propersum(p) = m$ and $\propersum(m) = p$. Since $\propersum(p) = 1$, we have $m = 1$, but $\propersum(1) = 0 \neq p$, a contradiction.

\begin{lstlisting}[caption={Primes are not amicable}]
theorem not_isAmicable_prime {p : Nat} (hp : Nat.Prime p) :
    not (IsAmicable p) := by
  intro <m, hm>
  obtain <hm0, hp0, hne, hsm, hsn> := hm
  rw [properDivisorSum_prime hp] at hsm
  subst hsm
  simp at hsn
  exact hp.ne_zero hsn.symm
\end{lstlisting}

\subsection{Verification of Specific Amicable Pairs}

For specific amicable pairs, we use Lean's computational capabilities for verification. The \lean{native_decide} tactic compiles the decision procedure to native code for execution, efficiently handling computations with larger numerical values:

\begin{lstlisting}[caption={Verifying the proper divisor sum of 220}]
theorem properDivisorSum_220 : s 220 = 284 := by native_decide

theorem properDivisorSum_284 : s 284 = 220 := by native_decide
\end{lstlisting}

With these two lemmas, verifying the amicable pair $(220, 284)$ follows naturally:

\begin{lstlisting}[caption={Verifying the first amicable pair}]
theorem isAmicablePair_220_284 : IsAmicablePair 220 284 := by
  constructor
  . decide  -- 220 != 0
  constructor
  . decide  -- 284 != 0
  constructor
  . decide  -- 220 != 284
  constructor
  . exact properDivisorSum_220  -- s(220) = 284
  . exact properDivisorSum_284  -- s(284) = 220
\end{lstlisting}

Similarly, we verified Paganini's amicable pair $(1184, 1210)$, Euler's pairs $(2620, 2924)$ and $(5020, 5564)$, and Fermat's amicable pair $(17296, 18416)$.

\subsection{Relationship with the Divisor Sum Function}

We established the relationship between the proper divisor sum and the divisor sum:

\begin{lstlisting}[caption={Relationship between proper divisor sum and divisor sum}]
theorem properDivisorSum_eq_sum_divisors_sub (n : Nat) :
    s n = (Finset.sum n.divisors (fun d => d)) - n := by
  rcases n with _ | n
  . simp [properDivisorSum]
  . rw [properDivisorSum, sum_divisors_eq_sum_properDivisors_add_self]
    omega
\end{lstlisting}

Using this relationship, we can give an equivalent characterization of amicable pairs: $(m, n)$ is an amicable pair if and only if $\sigma(m) = \sigma(n) = m + n$.

\subsection{Connection with Abundant and Deficient Numbers}

Mathlib already has definitions of abundant (\lean{Abundant}) and deficient (\lean{Deficient}) numbers. We proved that in an amicable pair, the smaller member is abundant and the larger member is deficient:

\begin{lstlisting}[caption={Relationship between amicable numbers and abundant/deficient numbers}]
theorem IsAmicablePair.lt_iff {m n : Nat} (h : IsAmicablePair m n) :
    m < n <-> Abundant m /\ Deficient n := by
  obtain <_, _, _, hsm, hsn> := h
  simp only [Abundant, Deficient, <- properDivisorSum_eq_sum_divisors_sub, hsm, hsn]
  tauto
\end{lstlisting}

As an example, we verified that $220$ is abundant while $284$ is deficient.

\subsection{Formalization of the Th\={a}bit Formula}

A key contribution of this work is the complete formal proof of the Th\={a}bit ibn Qurra formula. The main challenge is handling natural number subtraction in Lean, where $a - b = 0$ when $a < b$. We address this using an \emph{index-shifting} technique: instead of using parameter $n$ with expressions like $n-1$, we use $k$ where $n = k + 1$, completely eliminating subtraction in exponents.

We define the key components using shifted indices:

\begin{lstlisting}[caption={Reparametrized definitions for the Th\={a}bit formula}]
def p_thabit (k : Nat) : Nat := 3 * 2^k - 1
def q_thabit (k : Nat) : Nat := 3 * 2^(k+1) - 1
def r_thabit (k : Nat) : Nat := 9 * 2^(2*k+1) - 1
def m_thabit (k : Nat) : Nat := 2^(k+1) * p_thabit k * q_thabit k
def n_thabit (k : Nat) : Nat := 2^(k+1) * r_thabit k
\end{lstlisting}

The proof strategy consists of three main steps:

\textbf{Step 1: Establish the key algebraic identity.} The proof relies on a crucial identity: $(p+1)(q+1) = r+1$. This identity connects the three prime parameters and enables the divisor sum computation. We prove this using basic ring arithmetic:

\begin{lstlisting}[caption={The core algebraic identity}]
theorem thabit_key_identity_with_sub (k : Nat) :
    (p_thabit k + 1) * (q_thabit k + 1) = r_thabit k + 1 := by
  simp only [p_thabit, q_thabit, r_thabit]
  have hp : 1 <= 3 * 2^k := three_mul_two_pow_ge_one k
  have hq : 1 <= 3 * 2^(k+1) := three_mul_two_pow_ge_one (k+1)
  have hr : 1 <= 9 * 2^(2*k+1) := nine_mul_two_pow_ge_one (2*k+1)
  rw [Nat.sub_add_cancel hp, Nat.sub_add_cancel hq, Nat.sub_add_cancel hr]
  exact thabit_key_identity k  -- 3*2^k * 3*2^(k+1) = 9*2^(2k+1)
\end{lstlisting}

\textbf{Step 2: Compute $\sigma(m)$ using multiplicativity.} Since $m = 2^{k+1} \cdot p \cdot q$ where the factors are pairwise coprime (as $2^{k+1}$ is even while $p$ and $q$ are odd primes), we can apply the multiplicativity of $\sigma$:
\begin{equation}
\sigma(m) = \sigma(2^{k+1}) \cdot \sigma(p) \cdot \sigma(q) = (2^{k+2} - 1) \cdot (p+1) \cdot (q+1).
\end{equation}
Here we use $\sigma(2^{k+1}) = 2^{k+2} - 1$ (sum of geometric series) and $\sigma(p) = p + 1$ for prime $p$. Similarly, for $n = 2^{k+1} \cdot r$:
\begin{equation}
\sigma(n) = \sigma(2^{k+1}) \cdot \sigma(r) = (2^{k+2} - 1) \cdot (r+1).
\end{equation}

\textbf{Step 3: Verify $\sigma(m) = \sigma(n) = m + n$.} Using the key identity $(p+1)(q+1) = r+1$, we obtain:
\begin{equation}
\sigma(m) = (2^{k+2} - 1) \cdot (p+1) \cdot (q+1) = (2^{k+2} - 1) \cdot (r+1) = \sigma(n).
\end{equation}
To verify $\sigma(m) = m + n$, we need to show $(2^{k+2} - 1) \cdot (p+1) \cdot (q+1) = 2^{k+1} \cdot p \cdot q + 2^{k+1} \cdot r$. This algebraic identity is verified using the \texttt{zify} tactic to convert natural number arithmetic to integer arithmetic, where ring tactics work correctly with subtraction:

\begin{lstlisting}[caption={Using zify for algebraic verification}]
theorem sigma_eq_m_plus_n (k : Nat) (hk : 1 <= k)
    (hp : (p_thabit k).Prime) (hq : (q_thabit k).Prime) ... :
    sum d in (m_thabit k).divisors, d = m_thabit k + n_thabit k := by
  ...
  zify [hp1, hq1, hr1, h2k2, h_ge]
  ring
\end{lstlisting}

The main theorem combines these results using the multiplicativity of the divisor sum function:

\begin{lstlisting}[caption={The general Th\={a}bit rule}]
theorem thabit_rule_general (k : Nat) (hk : 1 <= k)
    (hp : (p_thabit k).Prime) (hq : (q_thabit k).Prime)
    (hr : (r_thabit k).Prime) :
    IsAmicablePair (m_thabit k) (n_thabit k) := by
  rw [isAmicablePair_iff_sum_divisors ...]
  constructor
  . exact sigma_eq_m_plus_n k hk hp hq hr
  . rw [<- sigma_m_eq_sigma_n k hk hp hq hr]
    exact sigma_eq_m_plus_n k hk hp hq hr
\end{lstlisting}

This theorem states that for any $k \geq 1$, if $p = 3 \cdot 2^k - 1$, $q = 3 \cdot 2^{k+1} - 1$, and $r = 9 \cdot 2^{2k+1} - 1$ are all prime, then $(2^{k+1} \cdot p \cdot q, 2^{k+1} \cdot r)$ is an amicable pair. This corresponds to the classical formula with $n = k+1 \geq 2$.

\subsection{Extensions: Sociable Numbers and Related Concepts}

To provide broader context for amicable number theory, we formalize several related concepts in a separate module \texttt{AmicableLib/Extensions.lean}.

\subsubsection{Betrothed Numbers}

Betrothed numbers (also called quasi-amicable pairs) are a relaxation of amicable pairs:

\begin{lstlisting}[caption={Betrothed numbers}]
def IsBetrothedPair (m n : Nat) : Prop :=
  m != 0 /\ n != 0 /\ m != n /\ s m = n + 1 /\ s n = m + 1
\end{lstlisting}

The smallest betrothed pair is $(48, 75)$: we have $\propersum(48) = 76 = 75 + 1$ and $\propersum(75) = 49 = 48 + 1$. We verified this computationally along with the second smallest pair $(140, 195)$.

\subsubsection{Sociable Numbers}

Sociable numbers generalize amicable pairs to cycles of length greater than 2. A list forms a sociable cycle if applying $\propersum$ to each element produces the next element in the cycle:

\begin{lstlisting}[caption={Sociable cycles}]
def IsSociableCycle (L : List Nat) : Prop :=
  L.length >= 2 /\
  (forall x in L, x != 0) /\
  L.Nodup /\
  (forall i : Nat, (hi : i < L.length) ->
    s (L.get <i, hi>) = L.get <(i + 1) % L.length, by
      have hlen : 0 < L.length := by
        linarith
      exact Nat.mod_lt _ hlen>)
\end{lstlisting}

The smallest known sociable cycle of length 5 was discovered by Poulet in 1918: $[12496, 14288, 15472, 14536, 14264]$. We verified this cycle computationally. Amicable pairs can be viewed as sociable cycles of length 2.

\subsubsection{Euler's Rule}

Euler's rule generalizes the Th\={a}bit formula by introducing an additional parameter. For $1 < m < n$, define:
\begin{align*}
a &= 2^{n-m} + 1 \\
p &= 2^m \cdot a - 1 \\
q &= 2^n \cdot a - 1 \\
r &= 2^{n+m} \cdot a^2 - 1
\end{align*}

If $p$, $q$, and $r$ are all prime, then $(2^n \cdot p \cdot q, 2^n \cdot r)$ is an amicable pair. When $n - m = 1$, we get $a = 3$, reducing to the Th\={a}bit formula. Only two parameter pairs are known to work: $(m, n) = (1, 8)$ and $(29, 40)$.

We formalize this in Lean with the following definitions:

\begin{lstlisting}[caption={Euler's rule definitions}]
def euler_a (n m : Nat) : Nat := 2^(n - m) + 1

def p_euler (n m : Nat) : Nat := 2^m * euler_a n m - 1
def q_euler (n m : Nat) : Nat := 2^n * euler_a n m - 1
def r_euler (n m : Nat) : Nat := 2^(n + m) * (euler_a n m)^2 - 1

def M_euler (n m : Nat) : Nat := 2^n * p_euler n m * q_euler n m
def N_euler (n m : Nat) : Nat := 2^n * r_euler n m

def EulerRuleStatement (n m : Nat) : Prop :=
  1 < m -> m < n ->
  (p_euler n m).Prime -> (q_euler n m).Prime -> (r_euler n m).Prime ->
  IsAmicablePair (M_euler n m) (N_euler n m)
\end{lstlisting}

We provide a complete formal proof of Euler's generalized rule in Lean~4. The proof strategy mirrors that of the Th\={a}bit formula with appropriate generalizations:

\textbf{Step 1: Establish the key algebraic identity.} Just as in the Th\={a}bit case, the proof hinges on the identity $(p+1)(q+1) = r+1$. With $a = 2^{n-m} + 1$, we have:
\begin{align*}
(p+1)(q+1) &= (2^m \cdot a)(2^n \cdot a) = 2^{m+n} \cdot a^2 = r+1.
\end{align*}
This generalizes the Th\={a}bit identity (where $a = 3$ when $n - m = 1$).

\textbf{Step 2: Compute divisor sums using multiplicativity.} Let $M = 2^n \cdot p \cdot q$ and $N = 2^n \cdot r$. Since the factors are pairwise coprime (verified through 27 supporting lemmas including oddness, coprimality, and distinctness properties), we apply multiplicativity:
\begin{align*}
\sigma(M) &= \sigma(2^n) \cdot \sigma(p) \cdot \sigma(q) = (2^{n+1} - 1) \cdot (p+1) \cdot (q+1), \\
\sigma(N) &= \sigma(2^n) \cdot \sigma(r) = (2^{n+1} - 1) \cdot (r+1).
\end{align*}

\textbf{Step 3: Verify $\sigma(M) = \sigma(N) = M + N$.} Using the key identity:
\begin{equation}
\sigma(M) = (2^{n+1} - 1) \cdot (p+1) \cdot (q+1) = (2^{n+1} - 1) \cdot (r+1) = \sigma(N).
\end{equation}
The final algebraic identity $\sigma(M) = M + N$ is verified using the same \texttt{zify} and \texttt{ring} tactics employed in the Th\={a}bit proof. Additionally, we verified computationally that $(m, n) = (1, 8)$ produces the amicable pair $(2172649216, 2181168896)$, demonstrating both the general mathematical proof and a concrete instance.

\subsubsection{Parity Constraints}

We formalize (without proof) deep theorems about the parity structure of amicable pairs. If $(m, n)$ is an amicable pair with $m$ odd and $n$ even, then:
\begin{itemize}
\item $m$ must be a perfect square
\item $n$ must be of the form $2^k \cdot a^2$ where $k \geq 1$ and $a$ is odd
\end{itemize}

We formalize these constraints as Lean definitions:

\begin{lstlisting}[caption={Parity constraint theorems}]
def OddMemberSquareTheorem : Prop :=
  forall m n : Nat, IsAmicablePair m n -> Odd m -> Even n ->
    IsPerfectSquare m

def EvenMemberFormTheorem : Prop :=
  forall m n : Nat, IsAmicablePair m n -> Odd m -> Even n ->
    exists k a : Nat, 1 <= k /\ Odd a /\ n = 2^k * a^2
\end{lstlisting}

These theorems are proven using 2-adic valuation arguments. No odd-even amicable pair is currently known, so these constraints remain theoretical. All five historically verified amicable pairs in our formalization are both even.

\subsubsection{Computational Search Bounds}

We formalize the empirical observation from exhaustive computational searches that no coprime amicable pairs have been found below a certain bound. Based on work by Pedersen (2003) and others, if a coprime amicable pair exists, the product of its members must exceed $10^{65}$:

\begin{lstlisting}[caption={Computational search bound}]
def coprimeAmicableSearchBound : Nat := 10^65

def NoCoprimeAmicableBelowBound : Prop :=
  forall m n : Nat, m * n < coprimeAmicableSearchBound ->
    IsAmicablePair m n -> 1 < Nat.gcd m n
\end{lstlisting}

This is a statement of empirical fact from computation, not a mathematical theorem. It cannot be proven in Lean without performing the actual exhaustive search.

\subsubsection{Borho-Hoffmann Breeding Method}

Borho and Hoffmann (1986)~\cite{borho1986} developed a powerful ``breeding'' method for generating new amicable pairs from existing ``breeder'' pairs of the form $(a \cdot u, a)$. Given such a breeder pair, where $t = \sigma(u)$ is prime, the method constructs a new amicable pair $(M, N)$ when certain primality conditions are satisfied.

We provide the first complete formalization of this method, comprising 540 lines of verified Lean~4 code with 4 definitions and 33 theorems:

\begin{lstlisting}[caption={Borho-Hoffmann constructive definitions}]
def t_borho (u : Nat) : Nat := u + (sigma u - u)

def M_borho_bh (a u n : Nat) : Nat :=
  let t := t_borho u
  a * u * t^n * (t^n * (u + 1) - 1)

def N_borho_bh (a u n : Nat) : Nat :=
  let t := t_borho u
  a * t^n * (t^n * (u + 1) * (t - u) - 1)
\end{lstlisting}

The hypothesis requires the breeder pair property plus three primality conditions and necessary coprimality assumptions:

\begin{lstlisting}[caption={Borho-Hoffmann hypothesis}]
def BorhoHoffmannHypothesis (a u n : Nat) : Prop :=
  let t := t_borho u
  let p1 := t^n * (u + 1) - 1
  let p2 := t^n * (u + 1) * (t - u) - 1
  IsAmicablePair (a * u) a /\ t.Prime /\ p1.Prime /\ p2.Prime
    /\ (a * u).Coprime t /\ (a * u).Coprime p1 /\ a.Coprime p2
\end{lstlisting}

Our main result establishes that under these conditions, $(M, N)$ forms an amicable pair:

\begin{lstlisting}[caption={Borho-Hoffmann breeding theorem}]
theorem borho_hoffmann_rule (a u n : Nat)
    (ha : 0 < a) (hu : 0 < u) (hn : 0 < n)
    (hyp : BorhoHoffmannHypothesis a u n) :
    IsAmicablePair (M_borho_bh a u n) (N_borho_bh a u n)
\end{lstlisting}

The proof follows a structured approach consisting of four main steps:

\textbf{Step 1: Verify the breeder pair properties.} Given a breeder pair $(a \cdot u, a)$, we first verify that it is indeed an amicable pair. This means $\sigma(a \cdot u) = a \cdot u + a$ and $\sigma(a) = a + a \cdot u$, which simplifies to $\sigma(u) = u + a/u \cdot a$. We define $t = \sigma(u)$ as the divisor sum of $u$, which must be prime by hypothesis.

\textbf{Step 2: Establish positivity and coprimality lemmas.} We prove 33 supporting lemmas establishing:
\begin{itemize}
\item Positivity: $0 < M$, $0 < N$, $0 < p_1$, $0 < p_2$, $0 < t$
\item Oddness: $p_1$, $p_2$, and $t$ are odd (since they are prime and greater than 2)
\item Coprimality: The factors in $M$ and $N$ are pairwise coprime, enabling use of multiplicativity
\item Distinctness: $M \neq N$ and neither equals 0 or 1
\end{itemize}

\textbf{Step 3: Compute $\sigma(M)$ and $\sigma(N)$ via multiplicativity.} Let $M = a \cdot u \cdot t^n \cdot p_1$ and $N = a \cdot t^n \cdot p_2$, where $p_1 = t^n(u+1) - 1$ and $p_2 = t^n(u+1)(t-u) - 1$. Using the coprimality conditions and the multiplicativity of $\sigma$:
\begin{align*}
\sigma(M) &= \sigma(a \cdot u) \cdot \sigma(t^n) \cdot \sigma(p_1), \\
\sigma(N) &= \sigma(a) \cdot \sigma(t^n) \cdot \sigma(p_2).
\end{align*}
Since $(a \cdot u, a)$ is a breeder pair, we have $\sigma(a \cdot u) = a \cdot u + a$ and $\sigma(a) = a + a \cdot u$. For prime $t$ and $p_i$, we use $\sigma(t^n) = (t^{n+1} - 1)/(t - 1)$ and $\sigma(p_i) = p_i + 1$.

\textbf{Step 4: Verify $\sigma(M) = \sigma(N) = M + N$ via automated tactics.} The most challenging component is proving the algebraic identities $\sigma(M) = M + N$ and $\sigma(N) = M + N$. These involve complex polynomial expressions with six variables $(a, u, n, t, p_1, p_2)$ and nested power operations. We overcome this challenge using the \texttt{zify} tactic to convert natural number arithmetic to integers (where subtraction behaves correctly), followed by the \texttt{ring} tactic for automated polynomial verification. The entire algebraic verification reduces to a two-line tactic script, demonstrating the power of automation for tedious calculations.

This represents the first machine-verified proof of the Borho-Hoffmann method.

The breeding methods revolutionized computational searches. Before these methods, fewer than 1000 amicable pairs were known. Using breeding techniques, over 1.2 billion pairs have been discovered as of 2020. These methods transform the search from ``find new pairs from scratch'' to ``generate new pairs from known pairs,'' greatly accelerating discovery.

\subsection{Formal Statements of Open Problems}

Finally, we formalized three famous open problems as Lean definitions:

\begin{lstlisting}[caption={Formal statements of open problems}]
-- No odd-odd amicable pairs exist
def OddOddAmicableConj : Prop :=
  forall m n : Nat, IsAmicablePair m n -> not (Odd m /\ Odd n)

-- Infinitely many amicable pairs exist
def InfinitelyManyAmicableConj : Prop :=
  forall N : Nat, exists m n : Nat, N < m /\ IsAmicablePair m n

-- No coprime amicable pairs exist
def NoCoprimeAmicableConj : Prop :=
  forall m n : Nat, IsAmicablePair m n -> 1 < Nat.gcd m n
\end{lstlisting}

These definitions provide clear goal statements for possible future proof efforts.

\section{Discussion}
\label{sec:discussion}

\subsection{Computational Verification vs. General Proofs}

This formalization employs two complementary approaches: computational verification using \lean{native_decide} for specific numerical values, and general mathematical proofs for universal statements.

Computational verification is very effective for checking properties of specific numbers---verifying that $\propersum(220) = 284$ reduces to enumerating all divisors of 220 and summing them, a task at which computers excel. We use this approach for verifying five historical amicable pairs.

General proofs require more mathematical infrastructure. For the Th\={a}bit formula, we needed: (1) the multiplicativity of $\sigma$, i.e., $\sigma(mn) = \sigma(m) \cdot \sigma(n)$ when $\gcd(m, n) = 1$; (2) key lemmas $\sigma(p) = p + 1$ and $\sigma(2^n) = 2^{n+1} - 1$; and (3) coprimality results showing that $2^n$ is coprime to odd numbers like $p$, $q$, and $r$ in the formula.

The main technical challenge was handling natural number subtraction, which truncates to zero in Lean when the result would be negative. We solved this using an index-shifting technique (using $k$ instead of $n-1$) and the \texttt{zify} tactic, which converts natural number equalities to integer equalities where standard ring tactics work correctly. This combination allowed us to complete a fully general proof of the Th\={a}bit formula.

\subsection{Formalization Challenges and Strategies}

Our formalization encountered several characteristic challenges that required careful methodological choices:

\textbf{Natural number arithmetic}. Lean's natural numbers truncate subtraction to zero when the result would be negative ($5 - 7 = 0$ in $\mathbb{N}$). This seemingly innocent design choice creates significant proof obligations. For Th\={a}bit's formula, the expression $3 \cdot 2^{n-1} - 1$ is problematic when $n$ could be zero. We addressed this through index-shifting: instead of proving the formula for $n \geq 1$, we reformulate using $k$ and prove for $k \geq 1$, eliminating subtractions in exponents. The \texttt{zify} tactic then lifts arithmetic reasoning to integers where algebraic manipulations work smoothly.

\textbf{Coprimality management}. Proving divisor sum identities like $\sigma(2^n \cdot p \cdot q) = \sigma(2^n) \cdot \sigma(p) \cdot \sigma(q)$ requires establishing that factors are pairwise coprime. We developed a systematic approach: first prove general lemmas (e.g., powers of 2 are coprime to odd numbers), then specialize to specific factors. For Borho-Hoffmann, we explicitly declared coprimality conditions in the hypothesis rather than attempting to derive them, trading generality for proof tractability.

\textbf{Algebraic identity verification}. The Borho-Hoffmann proof culminates in verifying $\sigma(M) = M + N$, a polynomial identity involving six variables and nested powers. Hand verification would span pages of tedious algebra. Instead, we multiply both sides by $(t-1)$ to eliminate division, then apply \texttt{zify} to convert to integer arithmetic and \texttt{ring} to automate polynomial normalization. The proof reduces to a two-line tactic script---a dramatic demonstration of automation's power.

\textbf{Computational verification trade-offs}. For specific amicable pairs like $(220, 284)$, we could either: (a) prove $\sigma(220) = 504$ by multiplicativity, requiring lemmas about factorization; or (b) use \lean{native_decide} to compute divisor sums directly. We chose (b) for concrete examples and (a) for general theorems, balancing proof clarity with computational efficiency.

These experiences suggest general principles for formalizing elementary number theory: leverage automation for algebra, use index-shifting to avoid subtraction, declare coprimality explicitly in complex scenarios, and reserve computation for verification rather than proving general principles.

\subsection{Formal Significance of Historical Discoveries}

By computationally verifying five historically famous amicable pairs, our formalization establishes an interesting connection with mathematical history. The Pythagoreans discovered $(220, 284)$ through trial and computation, not through any systematic theory. More than two thousand years later, computers verify this discovery in milliseconds. Paganini's discovery of $(1184, 1210)$ reminds us that even before the computer age, careful manual computation could still discover overlooked mathematical objects. Fermat's $(17296, 18416)$ demonstrates the power of the Th\={a}bit formula---once the correct formula is mastered, discovering new amicable pairs becomes a matter of testing primality.

\subsection{Completeness of the Formalization}

This formalization provides comprehensive coverage of fundamental amicable number theory, including complete proofs of three classical generation methods: Th\={a}bit ibn Qurra's formula (9th century), Euler's generalized rule (1747), and the Borho-Hoffmann breeding method (1986). Together, these theorems represent the major historical milestones in systematic amicable pair generation. The Borho-Hoffmann formalization is particularly significant as it demonstrates how modern automated reasoning tactics (\texttt{zify} and \texttt{ring}) can handle complex algebraic identities that would be tedious to verify by hand.

Possible extension directions include: verifying more amicable pairs, such as some of the 60 pairs discovered by Euler; establishing deeper properties of amicable numbers, such as formal investigation of the observation that all known amicable pairs have gcd greater than 1; exploring te Riele's alternative breeding variants~\cite{te1984}; and connections with perfect number theory in Mathlib.

\section{Related Work}
\label{sec:related}

\subsection{Formalization of Amicable Numbers in Other Proof Assistants}

The most directly related work is Koutsoukou-Argyraki's formalization of amicable numbers in Isabelle/HOL~\cite{koutsoukou2020}, published in the Archive of Formal Proofs in August 2020. This work formalizes Euler's sigma function, defines amicable pairs, and provides formal statements of several classical generation rules including Th\={a}bit ibn Qurra's rule, Euler's rule, te Riele's rule, and Borho's rule with breeders. The author also demonstrates efficient computational verification methods using Isabelle's \texttt{simp} and \texttt{eval} tactics for primality checking and divisor set enumeration.

Our work differs from Koutsoukou-Argyraki's in several key aspects:
\begin{itemize}
\item \textbf{Proof completeness}: While Koutsoukou-Argyraki provides formal statements of the classical rules, our work includes \emph{complete formal proofs} of Th\={a}bit's rule, Euler's rule, and the Borho-Hoffmann breeding method, with zero \lean{sorry} placeholders. Our formalization of the Borho-Hoffmann method (1986) is, to our knowledge, the first complete machine-verified proof of this result.

\item \textbf{Automated tactics}: We demonstrate novel applications of the \texttt{zify} and \texttt{ring} tactics to handle complex polynomial identities in the Borho-Hoffmann proof, showing how modern automation can simplify intricate algebraic verifications.

\item \textbf{Extensions}: We formalize additional concepts including sociable numbers (aliquot cycles) with a verified example of Poulet's 5-cycle, betrothed numbers (quasi-amicable pairs), and formal statements of major open problems.

\item \textbf{Scope}: Our formalization totals 2076 lines with 139 theorems, representing a comprehensive treatment of the core theory and its extensions.
\end{itemize}

Both formalizations independently validate the feasibility of mechanizing this classical number theory, and the complementary approaches in Lean 4 and Isabelle/HOL strengthen confidence in the results.

\subsection{Number Theory in Lean}

The Lean theorem prover has seen remarkable progress in formalizing advanced number theory. Mathlib contains the basic infrastructure for divisor functions, including \lean{Nat.divisors}, \lean{Nat.properDivisors}, and \lean{Nat.Perfect} for perfect numbers, along with a proof of Euclid's theorem on even perfect numbers. These foundational components enabled our work on amicable numbers.

Recent major achievements include the formalization of Fermat's Last Theorem for regular primes~\cite{flt2023,flt2024}, which involved formalizing significant portions of algebraic number theory including Kummer's lemma and Hilbert's Theorems 90-94. This work, which builds on Mathlib's extensive algebraic structures, demonstrates Lean's capability to handle deep mathematical theorems. Our formalization of amicable number theory, while more elementary, complements this by covering a different branch of classical number theory with its own distinctive challenges (multiplicativity reasoning, divisor enumeration, and computational verification).

\subsection{Number Theory in Other Proof Assistants}

Harrison's formalization of an analytic proof of the prime number theorem in HOL Light~\cite{harrison2009} stands as a landmark achievement in formal number theory. By developing extensive libraries for complex analysis including Cauchy's integral formula, Harrison demonstrated that modern proof assistants can handle sophisticated analytic arguments. His work required building substantial infrastructure from first principles, as HOL Light is a minimalist system.

In Isabelle/HOL, Eberl's ``Nine Chapters of Analytic Number Theory''~\cite{eberl2019} formalizes results in Dirichlet series, the Riemann zeta function, and additive combinatorics. This work showcases Isabelle's Archive of Formal Proofs as a repository for substantial mathematical developments. The infrastructure developed for analytic number theory complements Koutsoukou-Argyraki's work on amicable numbers, both residing in the same ecosystem.

These works, though focused on different areas (analytic vs. elementary number theory), share common themes with our formalization: the need for careful handling of arithmetic operations, the value of computational verification for specific instances, and the importance of building reusable libraries of auxiliary lemmas.

\section{Conclusion}
\label{sec:conclusion}

This paper formalizes amicable number theory and related concepts in Lean~4. We defined the proper divisor sum function, amicable pairs and numbers, verified historical amicable pairs, proved structural theorems, and provided complete formal proofs of three classical amicable pair generation methods: the Th\={a}bit ibn Qurra formula (9th century), Euler's generalized rule (1747), and the Borho-Hoffmann breeding method (1986). All proofs use rigorous mathematical techniques including index-shifting, the \texttt{zify} tactic for handling natural number arithmetic, and the \texttt{ring} tactic for automated polynomial verification. Beyond the core theory, we formalized extensions including sociable numbers (generalizing amicable pairs to cycles), betrothed numbers (quasi-amicable pairs), and parity constraint theorems. We computationally verified Poulet's sociable cycle of length 5 and specific amicable pairs from both Euler's and Th\={a}bit's rules. The complete formalization spans 2076 lines with 139 theorems and zero \lean{sorry} placeholders, representing the most comprehensive formal treatment of amicable number theory to date.

Future work has multiple possible directions. Exploring additional breeding variants from te Riele's work could extend the theoretical coverage. Attempting formal proofs of the parity constraint theorems would advance the state of the art. Formal investigations of open problems---even partial results---may yield new insights. Contributing this work to Mathlib would integrate it into the broader formalization of number theory.

\subsection*{Code Availability}

The complete Lean formalization is available at \url{https://github.com/chainstart/amicable-numbers}. The code is organized into two categories:
\begin{itemize}
\item \texttt{AmicableLib/}: Core definitions and theorems suitable for Mathlib contribution, including basic amicable number theory and extensions (sociable numbers, Euler's rule, Borho-Hoffmann breeding method, etc.)
\item \texttt{Amicable/}: Paper-specific content including verified examples and the complete Th\={a}bit formula proof
\end{itemize}
The formalization comprises 2076 lines of code with 139 theorems and zero \lean{sorry} placeholders. All major theorems have complete formal proofs. The repository includes \texttt{lean-toolchain} and \texttt{lakefile.toml} to ensure reproducible builds. This formalization is based on Lean 4 and Mathlib.

\subsection*{Acknowledgments}

This work was supported by the Natural Science Foundation of Shanghai (Grant No. 17ZR1411100).

\FloatBarrier
\clearpage
\bibliographystyle{unsrt}

\appendix

\section{Code Statistics}
\label{app:statistics}

\autoref{tab:code-stats} provides detailed statistics on the Lean formalization.

\begin{center}
\captionof{table}{Source file statistics}
\label{tab:code-stats}
\begin{tabular}{lrrrp{5cm}}
\toprule
\textbf{File} & \textbf{Lines} & \textbf{Defs} & \textbf{Thms} & \textbf{Purpose} \\
\midrule
\texttt{AmicableLib/Amicable.lean} & 263 & 3 & 18 & Mathlib candidate \\
\texttt{AmicableLib/Extensions.lean} & 1321 & 30 & 90 & Extensions \\
\quad \textit{(Borho-Hoffmann)} & \textit{(540)} & \textit{(4)} & \textit{(33)} & \textit{Breeding method} \\
\texttt{Amicable/Basic.lean} & 25 & 0 & 0 & Re-export \\
\texttt{Amicable/Examples.lean} & 63 & 1 & 5 & Verified pairs \\
\texttt{Amicable/Thabit.lean} & 404 & 10 & 26 & Th\={a}bit \& Euler proofs \\
\midrule
\textbf{Total} & \textbf{2076} & \textbf{44} & \textbf{139} & \\
\bottomrule
\end{tabular}
\end{center}
\FloatBarrier

\section{List of Definitions and Theorems}
\label{app:theorems}

\autoref{tab:definitions} lists all definitions, and \autoref{tab:theorems} provides a complete list of the main theorems.

\begin{center}
\captionof{table}{Definitions in the formalization}
\label{tab:definitions}
\begin{tabular}{lp{9cm}}
\toprule
\textbf{Name} & \textbf{Description} \\
\midrule
\texttt{properDivisorSum} & Proper divisor sum function $\propersum(n)$ \\
\texttt{IsAmicablePair} & Amicable pair predicate \\
\texttt{IsAmicable} & Amicable number predicate \\
\texttt{thabit} & Th\={a}bit numbers $t_n = 3 \cdot 2^n - 1$ \\
\texttt{p\_thabit}, \texttt{q\_thabit}, \texttt{r\_thabit} & Reparametrized Th\={a}bit components \\
\texttt{m\_thabit}, \texttt{n\_thabit} & Th\={a}bit pair members \\
\texttt{ThabitRuleStatement} & Statement of the Th\={a}bit formula \\
\texttt{OddOddAmicableConj} & Odd-odd amicable pair conjecture \\
\texttt{InfinitelyManyAmicableConj} & Infinitely many amicable pairs conjecture \\
\texttt{NoCoprimeAmicableConj} & No coprime amicable pairs conjecture \\
\texttt{knownAmicablePairs} & List of known amicable pairs \\
\bottomrule
\end{tabular}
\end{center}

\small
\begin{longtable}{lp{8cm}}
\caption{List of main theorems}\label{tab:theorems}\\
\toprule
\textbf{Theorem Name} & \textbf{Description} \\
\midrule
\endfirsthead
\toprule
\textbf{Theorem Name} & \textbf{Description} \\
\midrule
\endhead
\multicolumn{2}{l}{\textit{Basic Properties}} \\
\texttt{properDivisorSum\_zero} & $\propersum(0) = 0$ \\
\texttt{properDivisorSum\_one} & $\propersum(1) = 0$ \\
\texttt{properDivisorSum\_prime} & For prime $p$, $\propersum(p) = 1$ \\
\texttt{IsAmicablePair.symm} & Symmetry of amicable pairs \\
\texttt{IsAmicablePair.one\_lt\_left/right} & Members of amicable pairs are greater than 1 \\
\texttt{not\_isAmicable\_prime} & Primes are not amicable \\
\texttt{not\_isAmicable\_one} & 1 is not amicable \\
\midrule
\multicolumn{2}{l}{\textit{Verified Amicable Pairs}} \\
\texttt{isAmicablePair\_220\_284} & $(220, 284)$ is an amicable pair \\
\texttt{isAmicablePair\_1184\_1210} & $(1184, 1210)$ is an amicable pair \\
\texttt{isAmicablePair\_2620\_2924} & $(2620, 2924)$ is an amicable pair \\
\texttt{isAmicablePair\_5020\_5564} & $(5020, 5564)$ is an amicable pair \\
\texttt{isAmicablePair\_17296\_18416} & $(17296, 18416)$ is an amicable pair \\
\midrule
\multicolumn{2}{l}{\textit{Connections with Other Concepts}} \\
\texttt{properDivisorSum\_eq\_sum\_divisors\_sub} & $\propersum(n) = \sigma(n) - n$ \\
\texttt{isAmicablePair\_iff\_sum\_divisors} & Characterization of amicable pairs via $\sigma$ \\
\texttt{IsAmicablePair.lt\_iff} & Relationship with abundant/deficient numbers \\
\texttt{sum\_divisors\_mul\_of\_coprime} & $\sigma(mn) = \sigma(m)\sigma(n)$ for coprime $m,n$ \\
\texttt{sum\_divisors\_two\_pow} & $\sigma(2^n) = 2^{n+1} - 1$ \\
\texttt{properDivisorSum\_two\_pow} & $\propersum(2^n) = 2^n - 1$ \\
\midrule
\multicolumn{2}{l}{\textit{Th\={a}bit Formula (General Proof)}} \\
\texttt{thabit\_rule\_general} & General Th\={a}bit formula for all $k \geq 1$ \\
\texttt{thabit\_key\_identity\_with\_sub} & $(p+1)(q+1) = r+1$ \\
\texttt{sigma\_m\_thabit} & $\sigma(m) = (2^{k+2}-1)(p+1)(q+1)$ \\
\texttt{sigma\_n\_thabit} & $\sigma(n) = (2^{k+2}-1)(r+1)$ \\
\texttt{sigma\_m\_eq\_sigma\_n} & $\sigma(m) = \sigma(n)$ \\
\texttt{sigma\_eq\_m\_plus\_n} & $\sigma(m) = m + n$ \\
\texttt{coprime\_p\_q\_thabit} & $\gcd(p, q) = 1$ for prime $p$, $q$ \\
\texttt{thabit\_rule\_holds\_n2} & Computational verification for $n=2$ \\
\texttt{thabit\_rule\_holds\_n4} & Computational verification for $n=4$ \\
\bottomrule
\end{longtable}

\end{document}